\documentclass[a4paper,preprint,12pt]{elsarticle}
\usepackage{amsfonts,amsmath,amssymb,amsthm}
\usepackage{times}
\usepackage{xcolor}
\usepackage{graphicx}
\biboptions{numbers,sort&compress} 
\journal{arXiv.org}

\let\ts=\textstyle

\def\fracskip{\mskip 1mu \relax}
\def\nfrac#1#2{{\fracskip#1\fracskip\over\fracskip#2\fracskip}}
\def\tfrac#1#2{{\ts\nfrac{#1}{#2}}}

\let\frac=\nfrac
\def\pd#1#2{\frac{\partial#1}{\partial#2}}
\def\pdd#1#2#3{\ifx#2#3\pd{^2#1}{#2^2}\else\pd{^2#1}{#2\partial#3}\fi }
\definecolor{BrickRed}{rgb}{0.588,0.098,0.055}

\let\BL=\biggl \let\BR=\biggr
 
\def\arb{is an arbitrary constant}
\def\arbs{are arbitrary constants}

\def\Equation#1. {\medbreak{\bfseries\itshape{Equation\kern.3333em\relax#1.}}\enspace\ignorespaces }
\def\Solution#1. {\medbreak{\bfseries\itshape{Solution\kern.3333em\relax#1.}}\enspace\ignorespaces }
\def\Solutions#1. {\medbreak{\bfseries\itshape{Solutions\kern.3333em\relax#1.}}\enspace\ignorespaces }
\def\Problem#1 {\medbreak{\bfseries\itshape{Problem\kern.3333em\relax#1}}\enspace\ignorespaces }

\newtheoremstyle{remark}{\medskipamount}{\medskipamount}
  {\small\rmfamily}{\parindent}{\footnotesize\sffamily}{.}{.5em}{}
\theoremstyle{remark}

\def\href#1#2{%
  \textcolor{blue}{#2}}

\begin{document}
\begin{frontmatter}
\title{Exact solutions to homogeneous and quasi-homogeneous systems of nonlinear ODEs}
\author[ipm]{Andrei D. Polyanin}
  \ead{polyanin@ipmnet.ru}
\author[ipm]{Alexei I. Zhurov\corref{cor1}}
  \ead{zhurovai@cardiff.ac.uk}
  \cortext[cor1]{Corresponding author}
\address[ipm]{Ishlinsky Institute for Problems in Mechanics, Russian Academy of Sciences,\\
  101 Vernadsky Avenue, bldg~1, 119526 Moscow, Russia}

\begin{abstract}
This note considers fairly general quasi-homogeneous systems of first-order nonlinear
ODEs and homogeneous systems of second-order nonlinear ODEs that contain arbitrary
functions of several arguments. It presents several exact solutions to these systems in
terms of elementary functions.
\end{abstract}

\end{frontmatter}

\section{Brief introduction}

Systems of ordinary differential equations are a common object of study in various
scientific disciplines. Many exact solutions to such systems can be found in the
handbooks~\cite{kam,polzai2018}. The article \cite{cal2021} presented partial solutions
to a class of systems of first-order nonlinear ordinary differential equations with
homogeneous polynomial right-hand sides. The current note deals with more general
(than in \cite{cal2021}), quasi-homogeneous systems of first-order nonlinear ODEs as
well as homogeneous systems of second-order nonlinear ODEs. It presents several exact
solutions to these systems in terms of elementary functions.

\section{Quasi-homogeneous systems of nonlinear first-order ODEs}

Consider the following quasi-homogeneous systems of $N$ first-order ordinary differential
equations for the unknowns $x_1=x_1(t)$, \dots, $x_N=x_N(t)$:
\begin{equation}
x'_{n}  =x_n^{m_n+1}F_n\biggl(\frac{x_2^{m_2/m_1}}{x_1},
  \frac{x_3^{m_3/m_1}}{x_1},\dots,\frac{x_N^{m_N/m_1}}{x_1}\biggr),\qquad
n =1,\ldots,N,
\label{e1}
\end{equation}
where $F_n(\dots)$ are given arbitrary functions, $m_n$ are arbitrary constants, and
$N$ is an arbitrary positive integer.

System \eqref{e1} preserves its form under the transformation
\begin{equation*}
t=\lambda^{-m_1}\bar t,\quad \ x_{n} =\lambda^{m_1/m_n}\bar x_n,\quad \ n=1,\ldots,N,
\label{e2}
\end{equation*}
where $\lambda>0$ is an arbitrary constant.

Assuming that $m_n\not = 0$ ($n =1,\ldots,N$), we look for an exact solution to system
\eqref{e1} in the form
\begin{equation}
x_{n}(t) =a_n(1+Ct)^{-1/m_n},\qquad n=1,\ldots,N,
\label{e2}
\end{equation}
where $C$ is an arbitrary constant and $a_n=x_n(0)$ are constants
(initial values of the unknowns) to be determined.

The system of equations \eqref{e1} admits an exact solution of the form \eqref{e2}
with the constants~$a_n$ related by the algebraic (or transcendental) constraints
\begin{equation}
a_{n}^{m_n}m_nF_n\biggl(\frac{a_2^{m_2/m_1}}{a_1},\frac{a_3^{m_3/m_1}}{a_1},
\dots,\frac{a_N^{m_N/m_1}}{a_1}\biggr)+C=0,\qquad
n =1,\ldots,N.
\label{e3}
\end{equation}

\textit{Example 1}. For $m_1=\cdots=m_N=m$, the quasi-homogeneous system
\eqref{e1} simplifies to become a homogeneous system that can be represented as
\begin{equation}
x'_{n}  =x_n^{m+1}F_n\biggl(\frac{x_2}{x_1},\frac{x_3}{x_1},\dots,\frac{x_N}{x_1}\biggr),\qquad
n =1,\ldots,N,
\label{e4}
\end{equation}
where $F_n(\dots)$ are arbitrary functions. The system of ODEs \eqref{e4} admits an
exact solution of the form \eqref{e2} with $m_n=m$, where the constants $a_{n}$ are
related by the constraints
\begin{equation*}
a_{n}^{m}mF_n\biggl(\frac{a_2}{a_1},\frac{a_3}{a_1},\dots,\frac{a_N}{a_1}\biggr)+C=0,\qquad
n =1,\ldots,N.
\end{equation*}

The note \cite{eq2021} presents system \eqref{e4} and its solution for positive
integer~$m$.

In the degenerate case $m=0$, exact solutions to system \eqref{e4} can be sought in the
exponential form
\begin{equation*}
x_{n}(t) =a_n\exp(-Ct),\qquad n=1,\ldots,N,
\end{equation*}
where $C$ is an arbitrary constant. The constants $a_n$ and $C$ are related by the
constraints
\begin{equation*}
F_n\biggl(\frac{a_2}{a_1},\frac{a_3}{a_1},\dots,\frac{a_N}{a_1}\biggr)+C=0,\qquad
n =1,\ldots,N.
\end{equation*}

\textit{Example 2}.
The homogeneous system of ODEs \eqref{e4} can be represented in the equivalent form
\begin{equation}
x'_{n}  =x_1^{m+1}G_n\biggl(\frac{x_2}{x_1},\frac{x_3}{x_1},\dots,\frac{x_N}{x_1}\biggr),\qquad
n =1,\ldots,N,
\label{e5}
\end{equation}
where $G_n(\dots)=(x_n/x_1)^{m+1}F_n(\dots)$.
Suppose that the functions $G_n$ are all multivariate polynomials of degree $M=m+1$
such that
\begin{equation*}
G_n=\sum \alpha_{n\mu_2\ldots\mu_N}\biggl(\frac{x_2}{x_1}\biggr)^{\!\mu_2}\ldots
\biggl(\frac{x_N}{x_1}\biggr)^{\!\mu_N},\qquad
n =1,\ldots,N,
\end{equation*}
where $\alpha_{n\mu_2\ldots\mu_N}$ are some constants and $\mu_2$, \dots, $\mu_N$
are some nonnegative integers. Then system~\eqref{e5} becomes
\begin{equation*}
x'_{n}=\sum c_{n\mu_1\ldots\mu_N}x_1^{\mu_1}\ldots x_N^{\mu_N},\qquad
n =1,\ldots,N,
\end{equation*}
where $c_{n\mu_1\ldots\mu_N}$ are some constants and
$\mu_1=M-\mu_2-\cdots-\mu_N$. We see that $\mu_1+\cdots+\mu_N=M$. If $\mu_1$
is also a nonnegative integer, we obtain system (1) from \cite{cal2021}. It admits the
exact solution
\begin{equation*}
x_{n}(t) =a_n (1+Ct)^{1/(1-M)},\qquad n=1,\ldots,N,
\end{equation*}
where $C$ is an arbitrary parameter and the constants $a_n$ satisfy the algebraic
constraints
\begin{equation*}
\def\qquad{\ \ \ \ \ \ }
Ca_n=(1-M)\sum_{\mu_1+\cdots+\mu_N=M} c_{n\mu_{1}\ldots\mu_{N}}
a_1^{m_{1}}\ldots a_{N}^{m_{N}},\qquad
n =1,\ldots,N.
\end{equation*}

\section{Homogeneous systems of nonlinear second-order ODEs}

Now we look at homogeneous systems of second-order ODEs of the form
\begin{equation}
x''_{n}  =x_n^{m}(x'_n)^kF_n\biggl(\frac{x_2}{x_1},\frac{x_3}{x_1},\dots,
\frac{x_N}{x_1},\frac{x'_2}{x'_1},\frac{x'_3}{x'_1},\dots,\frac{x'_N}{x'_1}\biggr),\qquad
n =1,\ldots,N,
\label{e6}
\end{equation}
where $F_n(\dots)$ are arbitrary functions of their arguments, $m$ and $k$ \arbs,
and $N$ is an arbitrary positive integer.

Assuming all unknowns to be proportional, we look for a particular solution to system
\eqref{e6} in the special form
\begin{equation}
x_{n}=a_ny,\quad y=y(t),\quad n=1,\ldots,N,
\label{e7}
\end{equation}
where $a_n$ are constants to be determined. As a results, we arrive at the following
second-order ODE for~$y$:
\begin{equation}
y''=\lambda y^m(y')^k,
\label{e8}
\end{equation}
where $\lambda$ is an arbitrary constant, while $a_n$ satisfy the relations
\begin{equation*}
a_{n}^{m+k-1}F_n\biggl(\frac{a_2}{a_1},\frac{a_3}{a_1},\dots,\frac{a_N}{a_1},
\frac{a_2}{a_1},\frac{a_3}{a_1},\dots,\frac{a_N}{a_1}\biggr)=\lambda,\qquad
n =1,\ldots,N.
\end{equation*}
It is noteworthy that equation \eqref{e8} is solvable and its general solution
can be represented in implicit form.

For arbitrary $m$ and $k$ such that $m+k-1\not=0$, equation \eqref{e8} admits
the simple power-law particular solution
\begin{equation*}
y=A(1+Ct)^\sigma,\quad \sigma=\frac{k-2}{m+k-1},\quad
A=\BL[\frac{C^{2-k}(\sigma-1)}{\lambda\sigma^{k-1}}\BR]^{\ts\frac 1{m+k-1}},
\end{equation*}
where $C$ \arb.

For $m+k-1=0$, equation \eqref{e8} admits the exponential partial solution
\begin{equation*}
y=B\exp(Ct),\quad \lambda=C^{2-k},
\end{equation*}
where $B$ and $C$ \arbs.

Below are two examples of more complicated solutions to equation \eqref{e8} and system
\eqref{e6}.

\textit{Example 3}.
In the special case of $m=1$ and $k=0$, system \eqref{e6} becomes
\begin{equation*}
x''_{n}  =x_nF_n\biggl(\frac{x_2}{x_1},\frac{x_3}{x_1},\dots,\frac{x_N}{x_1},
\frac{x'_2}{x'_1},\frac{x'_3}{x'_1},\dots,\frac{x'_N}{x'_1}\biggr),\qquad
n =1,\ldots,N.
\end{equation*}
Its has two different exact solutions depending on the sign of~$\lambda$. These are
obtained from the second-order linear equation \eqref{e8} with $m=1$ and $k=0$:
\begin{equation}
\begin{aligned}
x_{n}&=a_n[C_1\exp(-\beta t)+C_2\exp(\beta t)] &&\text{if \ $\lambda=\beta^2>0$};\\
x_{n}&=a_n[C_1\cos(\beta t)+C_2\sin(\beta t)]  &&\text{if \ $\lambda=-\beta^2<0$},
\end{aligned}
\label{e9}
\end{equation}
where $C_1$ and $C_2$ \arbs. The constants $a_n$ satisfy the constraints
\begin{equation*}
F_n\biggl(\frac{a_2}{a_1},\frac{a_3}{a_1},\dots,\frac{a_N}{a_1},\frac{a_2}{a_1},
\frac{a_3}{a_1},\dots,\frac{a_N}{a_1}\biggr)=\pm \beta^2,\qquad
n =1,\ldots,N,
\end{equation*}
where the upper sign refers to the first group of solutions in~\eqref{e9}, while the
lower sign refers to the second group of solutions.

\textit{Example 4}. In the special case of $m=-3$ and $k=0$, system \eqref{e6}
becomes
\begin{equation*}
x''_{n}  =x_n^{-3}F_n\biggl(\frac{x_2}{x_1},\frac{x_3}{x_1},\dots,\frac{x_N}{x_1},
\frac{x'_2}{x'_1},\frac{x'_3}{x'_1},\dots,\frac{x'_N}{x'_1}\biggr),\qquad
n =1,\ldots,N.
\end{equation*}
An exact solution to this system can be found from equation \eqref{e8} with $m=-3$ and $k=0$.
It is given by
\begin{equation*}
x_{n}=a_n(C_1t^2+C_2t+C_3)^{1/2},\qquad
n =1,\ldots,N,
\end{equation*}
where $C_1$, $C_2$, and $C_3$ are constants of integration, which are related by
$C_1C_3-\tfrac14C_2^2=\lambda$. The constants~$a_n$ satisfy the constraints
\begin{equation*}
a_{n}^{-4}F_n\biggl(\frac{a_2}{a_1},\frac{a_3}{a_1},\dots,\frac{a_N}{a_1},
\frac{a_2}{a_1},\frac{a_3}{a_1},\dots,\frac{a_N}{a_1}\biggr)=C_1C_3-\tfrac14C_2^2,\qquad
n =1,\ldots,N.
\end{equation*}

\end{document}